\documentclass[twocolumn,aps,prb,10pt]{revtex4}
\usepackage{graphicx}
\usepackage{epstopdf}
\begin{document}
\title{Atomic-scale description \\of the paramagnetic susceptibility of non-magnetic Ba$_2$YMoO$_6$ 
}
\author{R. J. Radwanski}
\affiliation{Center of Solid State Physics, S$^{nt}$ Filip 5, 31-150 Krakow, Poland. email: sfradwan@cyf-kr.edu.pl}
\begin{abstract}
We succeeded in remarkably nice theoretical description
of the temperature dependence of the paramagnetic susceptibility of Ba$_2$YMoO$_6$ in the  whole temperature range as originating from the atomic-scale discrete electronic structure of Mo$^{5+}$ ions in the 4$d^1$ configuration resulting from crystal-field and spin-orbit interactions. A strong violation of the Curie-Weiss law and a nonmagnetic state of BYMo experimentally-observed down to 50~mK is caused by the dramatical reduction of the Mo-ion magnetic moment.
The quantitative reproduction of $\chi$(T) proves that almost all Mo ions are in the Mo$^{5+}$ state over the whole temperature range.
\end{abstract}
\maketitle
\section{Introduction}
Ba$_2$YMoO$_6$ (BYMo) is presently widely studied oxide with the 4$d$ open-shell ion \cite{1,2,3,4,5,6,7,8,9,10,11,12,13} exhibiting non-trivial temperature dependence of the paramagnetic susceptibility $\chi$(T), as is seen in Fig. 1, with an  dramatic decrease with the lowering temperature below 50~K of the $\chi^{-1}$(T) plot \cite{2,5,8}.
BYMo does not order magnetically down to 50 mK \cite{2,5,7} though $\chi$(T) in the 200-300 K region points to strong antiferromagnetic interactions as one would expect from $\theta_{CW}$ of -(160-200) K, see Fig. 1. \cite{1,2,7,8}. Authors of Ref. \cite{2} explained this persistent low-temperature non-magnetic state by amorphous arrangements of spin singlets. 
Quite recently Romhanyi {\it et al.} \cite{11} proposed, in contrast to a chiral spin-liquid model of Ref. \cite{9}, a spin-orbital
model with a dimer-singlet phase, composed of
random arrangement of spin-orbit dimers, without any
type of the long-range order. Such explanations sound much scientifically. They involve sophisticated mathematics but physical meaning of "spin singlets" or "dimer-singlet state" do not provide physical description of the anomalous behaviour of $\chi$(T) and physical mechanism of so large reduction of the effective magnetic moment and of the local magnetic moment itself.

In this paper we report the microscopic atomic-scale description of the temperature dependence of the paramagnetic susceptibility of Ba$_2$YMoO$_6$ as associated with the Mo$^{5+}$ ions in the crystal. In our description the on-site crystal-field (CEF) and spin-orbit (s-o) interactions are fundamentally important realising a general concept "From the atomic physics to the solid state physics" \cite{14}. 

\begin{figure}
\begin{center}
\vspace {-0.10 cm}
\includegraphics[width =7.9 cm]{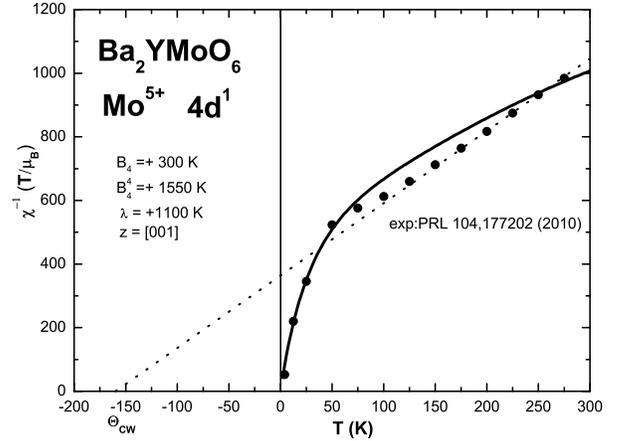} 
\end{center} 
\vspace {-0.3 cm}
\caption{Temperature dependence of the paramagnetic susceptibility of Ba$_2$YMoO$_6$ shown in the $\chi^{-1}$ plot where off-Curie-Weiss behaviour is more visible than in the $\chi$(T) plot, after Ref. \cite{2} (much similar data are in Refs \cite{5,8}). High-temperature dependence of the paramagnetic susceptibility of Ba$_2$YMoO$_6$ 
yields $\theta_{CW}$ of -160~K what would indicate on strong antiferromagnetic interactions, being in sharp contrast with a nonmagnetic state down to 50~mK, and with a quite substantial effective magnetic moment, about 1.5~$\mu_B$, at 300~K. The solid line is the result of this paper. See also Fig. 2.
}
\vspace {-0.3 cm}
\end{figure}
 
 \begin{table*}[h!tb]
\centering
\begin{tabular}{||c|c|ccccc|ccccc||}
\hline\hline
	E [K]	&	&+2				&-2				&		+1		&	-1				&	0 &S$_z$&L$_z$&J$_z$&m$_z$&m$_x$\\
\hline\hline
+22250	&1	&+0.727$\uparrow$	&+0.686$\uparrow$	&				&	+0.021$\downarrow$	& &+0.4996&+0.1158&+0.6154&+1.1161&-1.0294\\

37690 &10	&+0.686$\downarrow$&+0.727$\downarrow$&+0.021$\uparrow$	&	&		    	&-0.4996 &-0.1158&-0.6154&-1.1161&1.0294\\
\hline
+21650&5	&				&				&+0.037$\downarrow$	&				& 0.999$\uparrow$ &+0.4986&+0.0014&+0.5000&+0.9998&+1.0899\\
37090		&6	&		&			&				&+0.037$\uparrow	$& 0.999$\downarrow$ &-0.4986&-0.0014&-0.5000 &-0.9998 &-1.0899\\

\hline
-13460 &3	&+0.317$\downarrow$	&-0.325$\downarrow$ &+0.891$\uparrow$	&	& &+0.2942&+0.7843&+1.0785&+1.3733&-0.7844\\
	1980	&8	&-0.325$\uparrow$	&+0.317$\uparrow$	&		&+0.891$\downarrow$&  &-0.2942&-0.7843&-1.0785&-1.3733&+0.7844\\
\hline
-15000	&4	&			&				&0.999$\downarrow$&	 & -0.037$\uparrow$  &-0.4986&+0.9986&+0.5000 &+0.0002&+0.0886\\
440		&7	&				&				&	&0.999$\uparrow$	& -0.037$\downarrow$ &+0.4986&-0.9986&-0.5000 &-0.0002&-0.0886\\
\hline
-15440	&2	&+0.655$\downarrow$&-0.605$\downarrow$	&-0.453$\uparrow$	&	&			& -0.2946 &+0.3315&+0.0369&-0.2584&-0.2484\\
0		&9	&-0.605$\uparrow$	&+0.655$\uparrow$	&				&-0.453$\downarrow$			&  &+0.2946&-0.3315&-0.0369&+0.2584&+0.2484\\
\hline\hline
\end{tabular}
\vspace{0.2 cm}
\caption{Realised spin-orbital states, energies and their eigenfunctions ($|LSL_zS_z>$ space, with $L$=2 and $S$=1/2) of the Mo$^{5+}$ ion in Ba$_2$YMoO$_6$ calculated with the octahedral crystal-field parameter $B_4$= +300 K (10$Dq$=3.1~eV), an in-plane tetragonal-distortion  parameter $B_4^4$= +50~K (so $B_4^4$= +1550~K) and the spin-orbit coupling $\lambda_{s-o}$ = +1100~K. $L_z$ values as +2, -2, +1, -1, and 0 contributions are shown, whereas $S_z$ = +1/2 or -1/2 are shown by up and down arrows, respectively. In the first column the absolute and the relative energies are given. The second column is not informative - it shows a number of the row of the output results with eigenvalues and eigenfunctions from our home-made computer program. In last five columns the resulting magnetic characteristics from those eigenfunctions are shown. Note, a rather small $S_z$ and quite large values of $L_z$ in some cases and $J_z$ values much different from 3/2.  }
\label{tab:1}
\vspace{-0.3 cm}
\end{table*}
\section{Theoretical outline}
We have used the crystal-field approach developed already 50 years ago in a text-book by Abragam and Bleaney for paramagnetic impurities \cite{15}. Although it is a text book of 50 years old it only recently has started to be popular for description of oxides containing $d$ open-shell paramagnetic ions with strong s-o interactions in iridates and ruthenates \cite{16}. 
In approach, which we call as the Quantum Atomistic Solid-State Theory (QUASST), the magnetic and electronic properties of open-shell oxides are largely determined by single-ion electronic structure determined by crystal-field and spin-orbit interactions \cite{14}. CEF interactions result from charge interactions of the given paramagnetic ion in a crystal with its charge surroundings. A benchmark of the CEF+s-o-based QUASST is the integer valency of the involved transition-metal ion and the existence of the discrete low-energy electronic structure. As quite early applicability of such description to rare-earth and actinides compounds we can mention description of ErNi$_5$ and UPd$_2$Al$_3$ \cite{17}. For 3$d$-ion compounds we can mention description of FeBr$_2$ \cite{18} and LaCoO$_3$ \cite{19}. As ionic compound with the 3$d^1$ configuration we can mention BaVS$_3$, where we pointed out importance of the s-o interactions even in 3$d$ oxide manifested by substantial cancellation of the spin moment by the orbital moment \cite{20}. 

\section{Results and discusssion}
In this Letter we report on a found atomic-scale description of $\chi$(T) of BYMo in the whole temperature range, without a need for a $\chi_o$ correction, from 0.01 K to 300 K using only three physical parameters;
\begin{enumerate}
    \item octahedral crystal-field parameter $B_4^0$
 = +300~K, demanding $B_4^4$ = 5$\cdot B_4^0$ = 1500~K and yielding the largest energy scale of 10$Dq$ =~3.1 eV,
    \item  spin-orbit interactions with $\lambda_{s-o}$ = +1100 K yielding energy scale of 150 meV, and
    \item a small off-octahedral distortion, visible here in a change of the $B_4^4$ parameter to 1550~K, yielding a distortion energy of 40 meV.
\end{enumerate}

Ba$_2$YMoO$_6$ crystallises in the double perovskite structure (Fm\={3}m; SG 225).
In the double-perovskite structure the Mo$^{5+}$ (4$d^1$) ions are in the oxygen octahedral surroundings.

The usual single-ion Hamiltonian for the octahedral crystal field with a tetragonal distortion, completed by the s-o interaction term, takes a form in the Stevens notation \cite{14}
\begin{equation}  
\hat{H}= 
B_2^0 \hat{O}_2^0(L,L_z) + 
B_4^0 \hat{O}_4^0 (L,L_z)+ B_4^4 \hat{O}_4^4(L,L_z) + \lambda_{s-o} \vec{L} \cdot \vec{S}  
\end{equation}
where $\hat{O}_2^0$, $\hat{O}_4^0$ and $\hat{O}_4^4$ depend only on $L$ and $L_z$ (for the $^2D$ subterm relevant to the 4$d^1$ configuration of the Mo$^{5+}$ ion $L$=2 and $L_z$= 0, $\pm$1, $\pm$2) and where, for instance, $\hat{O}_2^0$ = 3$L_z^2$ - $L$($L$+1), and $S$=1/2. For the exactly octahedral local symmetry $B_4^4$= 5$\cdot B_4^0$. 
 
The effect of the octahedral CEF is well known - it splits 5 orbital states of the $^2D$ subterm ($L$=2, $S$=1/2) into the lower orbital triplet $t_{2g}$ and higher orbital doublet $e_g$ separated by 10$Dq$ (= 24$\cdot B_4^4$). 

\begin{figure}
\begin{center}
\vspace {-0.10 cm}
\includegraphics[width =8.50 cm]{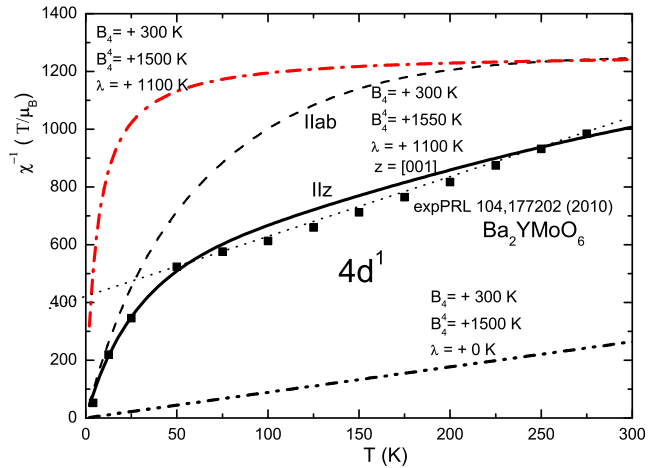} 
\end{center} 
\vspace {-0.4 cm}
\caption{Calculated temperature dependence of the paramagnetic susceptibility of Ba$_2$YMoO$_6$ associated with the atomic-scale electronic structure of the Mo$^{5+}$ (4$d^1$) ions (parameters are given in the figure, the realised eigenfunctions are collected in Table I) together with experimental data taken from Ref.\cite {2} (much similar data are in Refs \cite{5,8}). The dashed-point line (red) shows $\chi$(T) for exactly octahedral CEF (10$Dq$ = 3.1~eV with the spin-orbit coupling $\lambda$ = 
+95~meV) whereas the dashed-two-point line (blue) shows $\chi$(T) without the spin-orbit coupling. The solid and dashed lines show $\chi$(T) with small off-octahedral distortion with $\lambda$=~+95 meV for parallel (z) and perpendicular directions. 
} 
\vspace {-0.3 cm}
\end{figure}

 \begin{table*}[htb]
\centering
\begin{tabular}{||c|c|ccccc|cccc||}
\hline\hline
	E [K]	&	&+2				&-2				&		+1		&	-1				&	0 &S$_z$&L$_z$&J$_z$&m$_z$\\
\hline\hline
+21650	&1	&+0.728$\uparrow$	&+0.685$\uparrow$	&				&	+0.021$\downarrow$	& &+0.4995&+0.1196&+0.6192&+1.1199\\

36650 &10	&+0.685$\downarrow$&+0.728$\downarrow$&+0.021$\uparrow$	&	&		    	&-0.4995 &-0.1196&-0.6192&-1.1199\\
\hline
+21650&5	&				&				&+0.037$\downarrow$	&				& 0.999$\uparrow$ &+0.4986&+0.0014&+0.5000&+0.9998\\
36650		&6	&		&			&				&+0.037$\uparrow	$& 0.999$\downarrow$ &-0.4986&-0.0014&-0.5000 &-0.9998\\

\hline
-13300 &3	&+0.408$\downarrow$	&-0.408$\downarrow$ &+0.816$\uparrow$	&	& &+0.1671&+0.6667&+0.8333&+1.0000\\
	1700	&8	&-0.408$\uparrow$	&+0.408$\uparrow$	&		&+0.816$\downarrow$&  &-0.1671&-0.6667&-0.8333&-1.0000\\
\hline
-15000	&4	&			&				&0.999$\downarrow$&	 & -0.037$\uparrow$  &-0.4986&+0.9986&+0.5000 &+0.0002\\
0		&7	&				&				&	&0.999$\uparrow$	& -0.037$\downarrow$ &+0.4986&-0.9986&-0.5000 &-0.0002\\
\hline
-15000	&2	&+0.603$\downarrow$&-0.551$\downarrow$	&-0.577$\uparrow$	&	&			& -0.1671 &+0.4529&+0.2858&+0.1183\\
0		&9	&-0.551$\uparrow$	&+0.603$\uparrow$	&				&-0.557$\downarrow$			&  &+0.1671&-0.4529&-0.2858&-0.1183\\
\hline\hline
\end{tabular}
\vspace{-0.1 cm}
\caption{Realised spin-orbital states, energies and their eigenfunctions ($|LSL_zS_z>$ space, with $L$=2 and $S$=1/2) of the Mo$^{5+}$ ion in Ba$_2$YMoO$_6$ calculated with the octahedral crystal-field parameter $B_4$= +300 K (10$Dq$=~3.1~eV)
and the spin-orbit coupling $\lambda_{s-o}$ = +1100~K. $L_z$ values as +2, -2, +1, -1, and 0 contributions are shown, whereas $S_z$ = +1/2 or -1/2 are shown by up and down arrows, respectively. In the first column the absolute and the relative energies are given. The second column is not informative - it shows a number of the row of the output results with eigenvalues and eigenfunctions from our home-made computer program. In last five columns the resulting magnetic characteristics from those eigenfunctions are shown. }
\label{tab:1}
\vspace{-0.3 cm}
\end{table*}

We think that we have found a nice reproduction of $\chi$(T) of Ba$_2$YMoO$_6$, see Fig. 1. There $\chi^{-1}$(T) is presented, because an off-Curie-Weiss dependence is more visible than in the $\chi$(T) plot. By introducing a small tetragonal distortion, by changing only $B_4^4$ from 1500 K to 1550 K, we have got an abrupt lowering of the $\chi^{-1}$(T) (or an abrupt increase of $\chi$(T)) below T=~50 K exactly as is experimentally observed. Simultaneously our calculations reproduce large lowering of the effective magnetic moment and reproduce the observed Curie-Weiss temperature $\theta_{CW}$ of -(160-200~K). In our calculations large value of $\theta_{CW}$ is an effect of crystal-field and s-o interactions, not at all related to antiferromagnetic interactions. The reproduction of the absolute value of $\chi$(T) proves that all Mo atoms contribute equally to the macroscopic susceptibility and that all Mo atoms occur in BYMo as the Mo$^{5+}$ ions in the whole temperature range.

Fig. 2 illustrates influence of the s-o and off-octahedral CEf interactions on temperature dependence of $\chi$(T) (of course, the change of $\chi$(T) is due to the changing of the energies of 5 orbitals and their eigenfunctions, see Table I). In the absence of s-o interactions (then 3 $t_{2g}$ orbitals are degenerated) $\chi$(T) follows exactly the Curie law with the effective moment of 2.26 $\mu_B$ (dashed-two-point line in Fig. 2). The s-o coupling in the presence of the octahedral CEF removes 3-fold orbital degeneracy for a lower doublet and  higher by 1700 K orbital singlet, as is shown in Table II, where also their eigenfunctions are collected. In this case the resulting $\chi$(T) down to T= 50~K looks like a Pauli susceptibility (the dashed-point line on the top of Fig. 2) having, however, so huge effective moment as 6.0 $\mu_B$. The inferred distortion $B_4^4$, which could be produced in the reality, for instance, by a slight shortening in-plane oxygen distances, causes anisotropy of the local susceptibility (the experimentally realised $\chi$(T) direction is along the larger susceptibility shown in Fig. 2 as the solid line, which well follows the experimental points read from Ref. \cite{2}) and removes double orbital degeneracy yielding three Kramers-doublet states at energies 0, 38 and 170 meV, see Table I. 

The Kramers-doublet ground state is characterised by largely reduced both spin (s$_z$=~$\pm$0.295) and magnetic moment (m$_z$=~$\pm$0.258~$\mu_B$). One sees, that the spin is reduced by the factor of 2, whereas the magnetic moment much more, by the factor of 4. Such large reduction of the magnetic moment is caused by a substantial orbital moment of m$_z$=~$\pm$0.33~$\mu_B$). This ground-state magnetic moment largely determines the lowest-temperature susceptibility yielding in the temperature range up to 25 K the Curie-Weiss law with $p_eff$=~0.574~$\mu_B$. This value is in perfect agreement with the experimental value of $p_eff$=~0.57~$\mu_B$ \cite{8}.  

Table I collects the spin-orbital states, energies and their eigenfunctions, calculated within the $|LSL_zS_z>$ space, with $L$=2 and $S$=1/2, relevant to the Mo$^{5+}$ ion in Ba$_2$YMoO$_6$ under the action of the octahedral crystal field, $B_4^0$= +300 K and $B_4^4$= +1500~K, of the relativistic on-site spin-orbit coupling $\lambda_{s-o}$ = +1100~K and a small tetragonal distortion obtained to be realised by extra $B_4^4$ = +50~K (so, finally $B_4^4$ = +1550~K). An occurrence of a local distortion of MoO$_6$ octahedra, despite of the overall cubic symmetry, has been also infreed fro infrared transmission spectroscopy experiments \cite{8}. Table II collects the spin-orbital states and their characteristics for the purely octahedral crystal field. Table II is presented for noticing the effect of the tetragonal distortion, if compared with Table I. One can notice a substantial increase of the local magnetic moment and the spin of the CEF+s-o ground state by the tetragonal distortion, to m$_z$=~$\pm$0.258~$\mu_B$ and s$_z$=~$\pm$0.295 from m$_z$=~$\pm$0.118~$\mu_B$ and s$_z$=~$\pm$0.167. Thus our calculations prove that distortions usually help the formation of the magnetic order.

Inspecting Table I one notices that each state has large orbital moment, which only partly cancels the spin moment (which on the other side can take itself quite small values). The ground state moment amounts to 0.25 $\mu_B$, but in both directions. Interesting is the diminishing anisotropy of the susceptibility when temperature approaches zero temperature - it is responsible for so persistent nonmagnetic state of BYMo or, in other words, for formation of an amorphous spin state of atomic-scale Mo ions, because the moments cannot establish one preferred direction. However, due to that the Kramers degeneracy of the ground state has not been yet removed we expect enormous low-temperature specific heat resembling heavy-fermion phenomena in anomalous Ce, U and Yb intermetallics. 
In both of these type compounds the intra-atomic relativistic s-o coupling plays 
the fundamental role for the final determination of the shape of the ground-state eigenfunctions with sophisticated values of the spin and orbital moments. The relativistic s-o interactions has allowed for the revealing of the orbital magnetic moment in NiO so large as 0.54~$\mu_B$ (20\% of the total moment) \cite{21} and as 0.80~$\mu_B$ associated with the Fe$^{2+}$ ions in FeBr$_2$ \cite{18}.


Although we have got a nice reproduction of $\chi$(T) in the whole temperature range with revealing the origin of the dramatic reduction of the local Mo moment 
with approaching the lowest temperatures we would like to mention the experimental specific heat of BYMo, Fig. 4 of Ref. \cite{2}, which has revealed a Schottky-like peak with the heat maximum at about 80~K.  Assuming that this peak is really the Schottky-like one one derives the existence of the local state at about 200~K.  In our case we have got the first excited orbital Kramers doublet at 440~K. At present, we did not succeed to get local parameters which would recoincile the $\chi$(T) and c(T) dependences but we think that within our approach it will be possible by introducing lower symmetry CEF parameters. If one is at this moment not convinced about our present contribution to the theoretical description of $\chi$(T) experiments on BYMo he is welcome to remember that all this discussion is going within 20~meV energy, whereas up to now the electronic structure has been discussed in the best case in a 0.3~eV-energy scale.      
\section{Conclusions}
We succeeded in remarkably nice theoretical description
of the temperature dependence of the paramagnetic susceptibility of Ba$_2$YMoO$_6$ in the  whole temperature range as originating from the atomic-scale discrete electronic structure of Mo$^{5+}$ ions in the 4$d^1$ configuration resulting from crystal-field and spin-orbit interactions. A strong violation of the Curie-Weiss law and a nonmagnetic state experimentally observed down to 50~mK, despite of s=~1/2 spins, results from the on-site CEF + spin-orbit interactions. The relativistic s-o and crystal-field interactions cause the dramatic decrease of the spin and magnetic moment of the Mo$^{5+}$ (4$d^1$) ions.
Our results indicate that BYMo should be described as a spin-glass system with incoherent intersite local distortions in which exchange interactions become too weak to produce a magnetic order above 50~mK due to the dramatic lowering of the Mo-ion magnetic moment.  
We treat our single-valence description, with all Mo ions being in the Mo$^{5+}$ state, as much physically superior over a description with many mixed Mo-ion valences \cite{13}.

\vspace{-0.0cm}  
 email: sfradwan@cyf-kr.edu.pl; www.actaphysica.eu

\end{document}